\documentclass[letterpaper,12pt]{article}

\pdfoutput=1
\usepackage{alltt}
\usepackage{listings}
\usepackage{framed}
\usepackage{graphicx}
\usepackage{epsfig}
\usepackage{ulem}
\usepackage{fullpage}
\usepackage{subfig}
\usepackage{url}
\usepackage{comment}

\title{Dissecting a Small InfiniBand Application Using the Verbs API}

\author{Gregory Kerr\thanks{This work was partially supported by the
National Science Foundation under Grant OCI-0960978.} \\ College of
Computer and Information Science \\ Northeastern University \\ Boston,
MA \\ kerrg@ccs.neu.edu }

\date{}

\begin{document}
\maketitle

\begin{abstract}
InfiniBand is a switched fabric interconnect. The InfiniBand specification
does not define an API. However the OFED package, libibverbs, has become
the default API on Linux and Solaris systems.  Sparse documentation exists
for the verbs API. The simplest InfiniBand program provided by OFED,
{\tt ibv\_rc\_pingpong}, is about 800 lines long.  The semantics of using
the verbs API for this program is not obvious to the first time reader.
This paper will dissect the {\tt ibv\_rc\_pingpong} program in an attempt
to make clear to users how to interact with verbs. This work was motivated
by an ongoing project to include direct InfiniBand support for the DMTCP
checkpointing package~\cite{DMTCP}.
\end{abstract}

\section{Introduction}

The program {\tt ibv\_rc\_pingpong} can be found at
openfabrics.org\footnote{http://www.openfabrics.org/downloads/OFED/},
under the {\tt ``examples/''} directory of the OFED tarball. The source code
used for this document is from version~1.1.4. The {\tt ibv\_rc\_pingpong}
program
sets up a connection between two nodes running InfiniBand adapters
and transfers data.  Let's begin by looking at the program in action. In
this paper, I will refer to two nodes: {\tt client} and {\tt server}. There are
various command line flags that may be set when running the program. It is important
to note that the information contained within this document is based
on the assumption that the program has been run with no command line
flags configured. Configuring these flags will alter much of the program's behavior.

\begin{figure}
\begin{verbatim}
[user@server]$ ibv_rc_pingpong
  local address:  LID 0x0008, QPN 0x580048, PSN 0x2a166f, GID ::
  remote address: LID 0x0003, QPN 0x580048, PSN 0x5c3f21, GID ::
8192000 bytes in 0.01 seconds = 5167.64 Mbit/sec
1000 iters in 0.01 seconds = 12.68 usec/iter
\end{verbatim}

\begin{verbatim}
[user@client]$ ibv_rc_pingpong server
  local address:  LID 0x0003, QPN 0x580048, PSN 0x5c3f21, GID ::
  remote address: LID 0x0008, QPN 0x580048, PSN 0x2a166f, GID ::
8192000 bytes in 0.01 seconds = 5217.83 Mbit/sec
1000 iters in 0.01 seconds = 12.56 usec/iter
\end{verbatim}
\end{figure}

Since both nodes run the same executable, the ``client'' is the instance
that is launched with a hostname as an argument. The LID, QPN, and PSN
will be explained later.

Before we delve into the actual code, please look at a list of all
verbs API functions which will be used for our purposes.  I encourage
the reader to pause and read the man page for each of these.
\begin{figure}
\caption{manpage entries for the verbs API}
\begin{verbatim}
ibv_get_device_list (3), ibv_open_device (3), ibv_alloc_pd (3),
ibv_reg_mr (3), ibv_create_cq (3), ibv_create_qp (3), ibv_modify_qp (3),
ibv_post_recv (3), ibv_post_send (3), ibv_ack_cq_events (3)
\end{verbatim}
\end{figure}

\section{Layers}

\begin{figure}
\centering
 \subfloat[Software Stack Layers]{\includegraphics[width=0.6\textwidth,keepaspectratio=true]{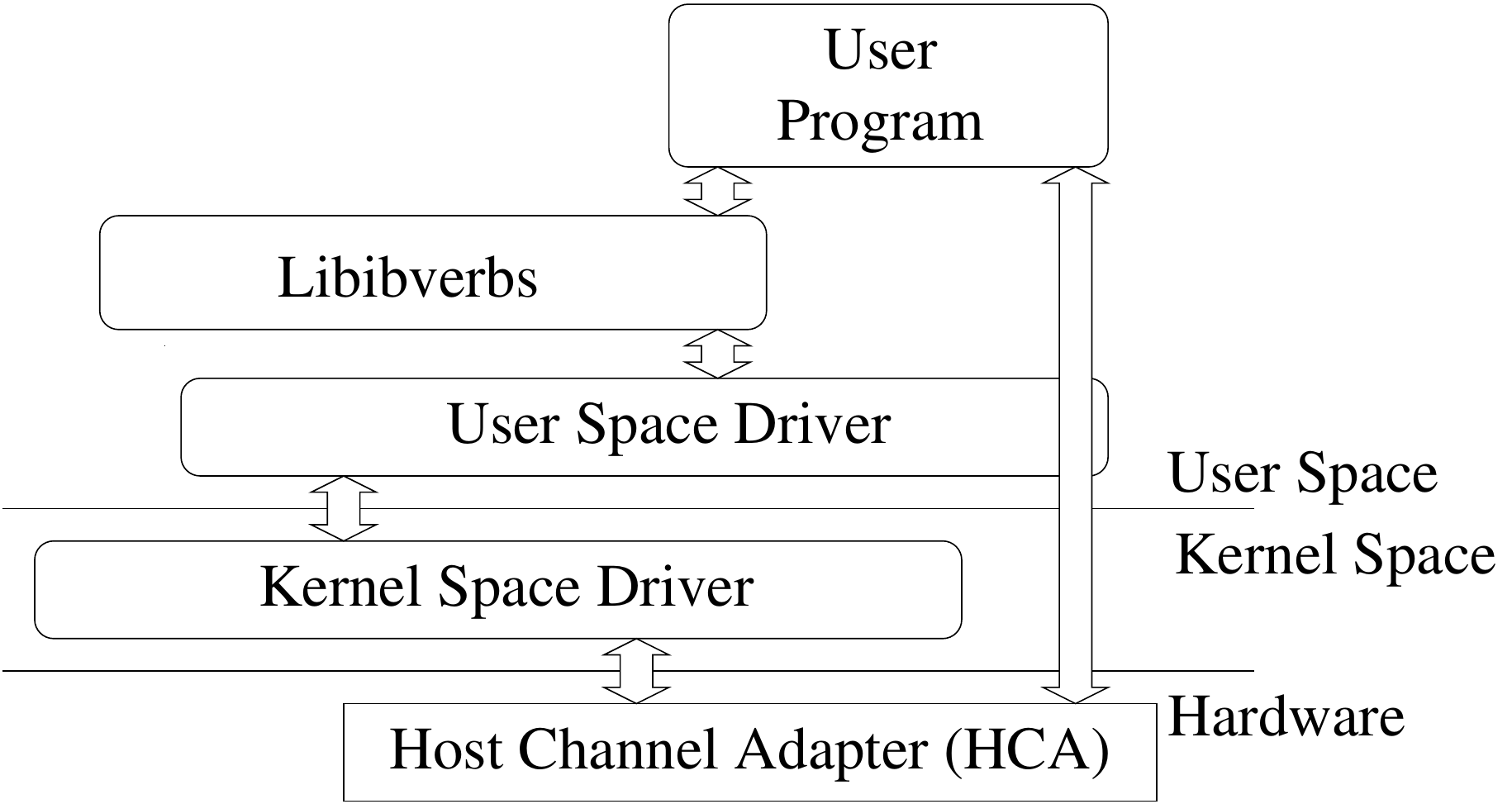}
 \label{fig:Layers}}
 \subfloat[Direct Memory Access]{\includegraphics[width=0.4\textwidth,keepaspectratio=true]{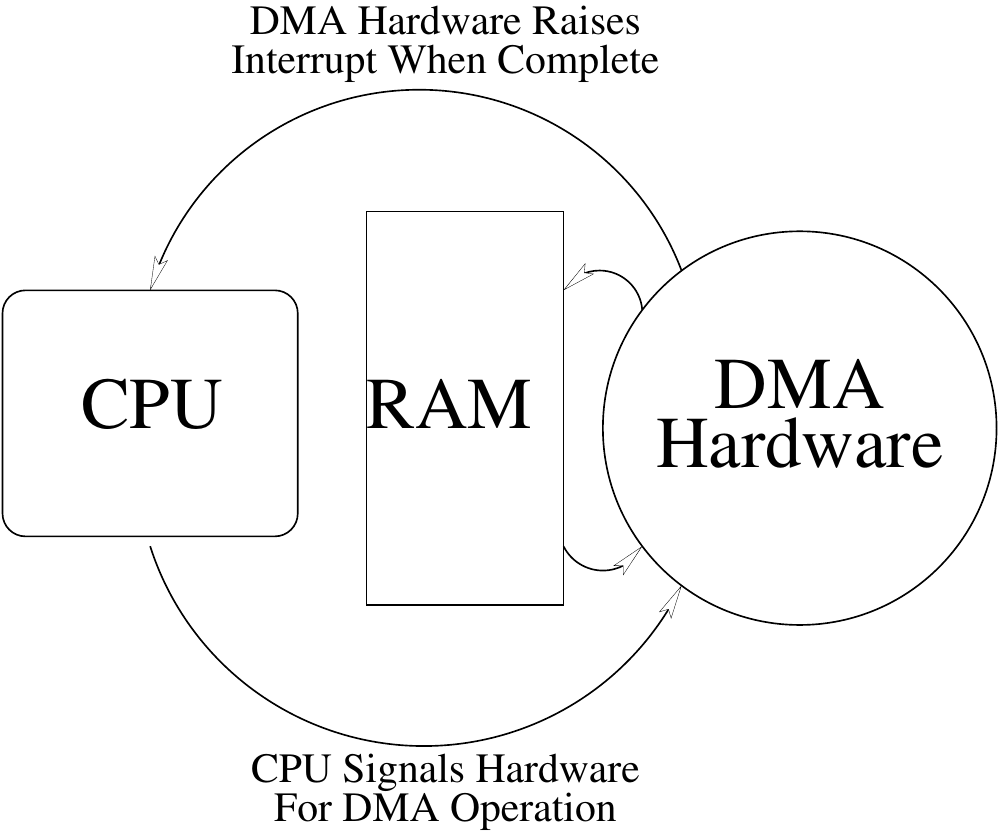}
\label{fig:dmaDiag}}
\caption{Layers and DMA}
\end{figure}

There are multiple drivers, existing in kernel and
userspace, involved in a connection. See Figure~\ref{fig:Layers}. To explain it simply,
much of the connection setup work goes through the kernel driver, as
speed is not a critical concern in that area.

The user space drivers are involved in function calls such as
{\tt ibv\_post\_send} and {\tt ibv\_post\_recv}. Instead of going through kernel
space, they interact directly with the hardware by writing to a segment of
mapped memory. Avoiding kernel traps is one way to decrease the overall
latency of each operation.

\section{Remote Direct Memory Access}

One of the key concepts in
InfiniBand is Remote Direct Memory Access (RDMA). This allows a node
to directly access the memory of another node on the subnet, without
involving the remote CPU or software layers.

Remember the key concepts of Direct Memory Access (DMA) as illustrated
by Figure~\ref{fig:dmaDiag}.

In the DMA, the CPU sends a command to the hardware to begin a DMA
operation. When the operation finishes, the DMA hardware raises an
interrupt with the CPU, signaling completion. The RDMA concept used
in InfiniBand is similar to DMA, except with two nodes accessing each
other's memory; one node is the sender and one is the receiver.

\begin{figure}[tbp]
\centering
 \includegraphics[width=400pt,keepaspectratio=true]{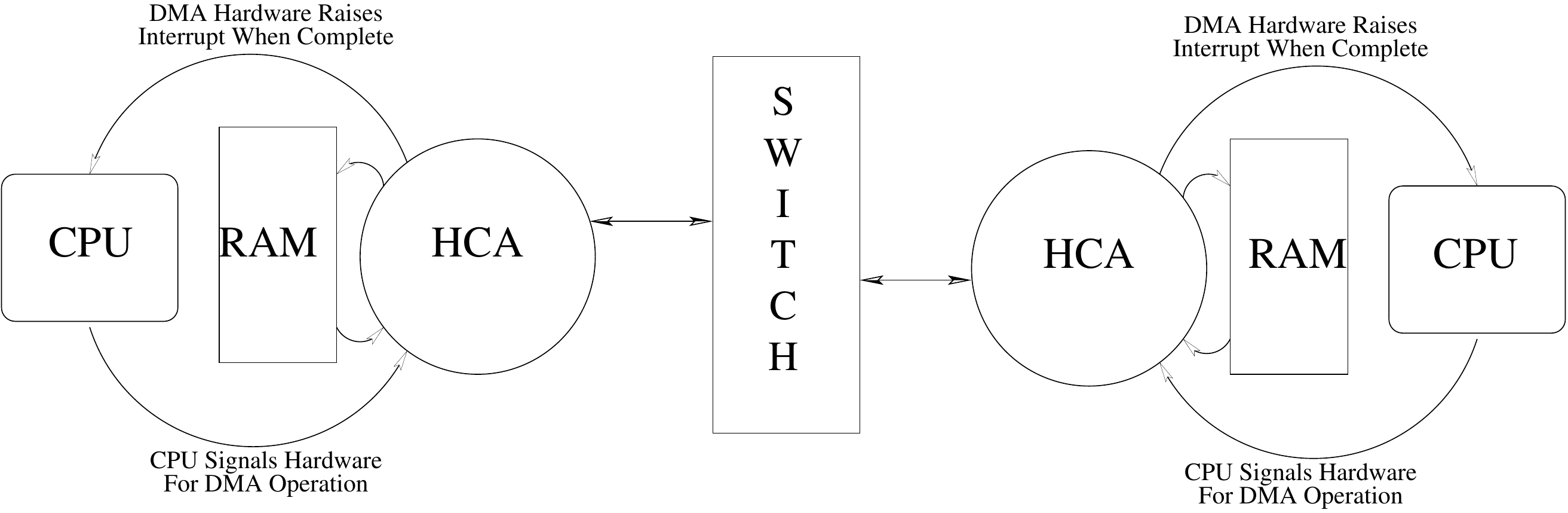}
\caption{InfiniBand Remote Direct Memory Acess}
\label{fig:rdmaDiag}
\end{figure}

Figure 2 illustrates an InfiniBand connection. In this case the DMA
Hardware is the Host Channel Adapter (HCA), and the two HCAs are
connected, through a switch, to each other. The HCA is InfiniBand's
version of a network card; it is the hardware local to each node that
facilitates communications. This allows an HCA on one node to use another
HCA to perform DMA operations on a remote node.

\section{Overview}

The {\tt ibv\_rc\_pingpong} program does the following.

\begin{enumerate}
\item Reserves memory from the operating system for sending and receiving data
\item Allocates resources from the verbs API
\item Uses a TCP socket to exchange InfiniBand connection information
\item Creates a connection between two InfiniBand ports
\item Transfers data over the connection
\item Acknowledges the successful completion of the transfer
\end{enumerate}

\section{Data Transfer Modes} The InfiniBand specification states
four different connection types: Reliable Connection (RC), Unreliable
Connection (UC), Reliable Datagram (RD), Unreliable Datagram (UD).
This program, {\tt ibv\_rc\_pingpong} uses a simple RC model.  RD is not
supported by current hardware.

The difference between reliable and unreliable is predictable -- in a
reliable connection data is transferred in order and guaranteed to arrive.
In an unreliable connection neither of those guarantees is made.

A connection type is an association strictly between two hosts. In a
datagram, a host is free to communicate with any other host on the subnet.

\section{Queue Based Model} The InfiniBand hardware processes requests
from the client software through requests, which are placed into queues.
To send messages between nodes, each node must have at minimum three
queues: a Send Queue (SQ), Receive Queue (RQ), and Completion Queue (CQ).

In a reliable connection, used in the {\tt ibv\_rc\_pingpong} program, queue
pairs on two distinct hosts compromise an end-to-end context. They send
messages to each other, and only each other. This paper restricts itself
to this mode.

The queues themselves exist on the HCA. However the libibverbs will
return to the user a data structure which corresponds with the QP.
While the library will create the QP, the user assumes the
responsibility of ``connecting'' the QP with the remote node. This is
generally done by opening an out-of-band socket connection, trading the
identification numbers for the queues, and then updating the hardware
with the information.

More recently, {\tt librdma\_cm} (an OFED library for connection management) allows
a user to create and connect QPs through library calls reminiscent of
POSIX sockets. Those calls are outside the scope of this document.

\subsection{Posting Work Requests to Queues} To send and receive data
in the InfiniBand connection (end-to-end context), work requests, which
become Work Queue Entries (WQE, pronounced ``wookie'') are posted to the
appropriate queue. These work
requests point to lists of scatter/gather elements (each element has an
address and size associated with it). This is a means of writing to and
reading from buffers which are non-contiguous in memory.

The memory buffers must be registered with the hardware; that process is
explained later.  \textbf{Memory buffers must be posted to the receive
queue before the remote host can post any sends}.  The {\tt ibv\_rc\_pingpong}
program posts numerous buffers to the receive queue at the beginning of
execution, and then repopulates the queue as necessary. A receive queue
entry is processed when the remote host posts a send operation.

When the hardware processes the work request, a Completion Queue Entry
(CQE, pronounced ``cookie'') is placed on the CQ.  There is a sample of
code showing how to handle completion events in {\tt ibv\_ack\_cq\_events (3)}.

\section{Connecting the Calls} The table below which the
function calls used in {\tt ibv\_rc\_pingpong} to create a connection, and
the order in which they are called.

\begin{center}
  \begin{tabular}{  r  l  }
    \hline
    struct ibv\_device ** & ibv\_get\_device\_list(int * num\_devices); \\ \hline
    struct ibv\_context * & ibv\_open\_device(struct ibv\_device * device); \\ \hline
    & /* protection domain */  \\
    struct ibv\_pd * & ibv\_alloc\_pd(struct ibv\_context * ctx); \\ \hline
    & /* memory region */ \\
    struct ibv\_mr * & ibv\_reg\_mr(struct ibv\_pd * pd, void * addr, size\_t length, \\
     &  enum ibv\_access\_flags access); \\ \hline
    & /* completion queue */ \\
    struct ibv\_cq * & ibv\_create\_cq(struct ibv\_context * context, int cqe, \\
& void * cq\_context,
     struct ibv\_comp\_channel channel, \\
& int comp\_vector); \\ \hline
    & /* queue pair */ \\
    struct ibv\_qp * & ibv\_create\_qp(struct ibv\_pd * pd,
    \\ & struct ibv\_qp\_init\_attr * qp\_init\_attr); \\ \hline
    int & ibv\_modify\_qp(struct ibv\_qp * qp, struct ibv\_qp\_attr * attr, \\
     & int attr\_mask); \\ \hline
  \end{tabular}
\end{center}

This table introduces the resources which are allocated in
the process of creating a connection. These resources will be
explained in detail later.

\section{Allocating Resources}

\subsection{Creating a Context} The first function call to the verbs API
made by the {\tt ibv\_rc\_pingpong} source code is here:

\lstset{language=C, caption=,basicstyle=\footnotesize}
\begin{lstlisting}
619     dev_list = ibv_get_device_list(NULL);
\end{lstlisting}

As the man page states, this function returns a list of available HCAs.

The argument to the function is an optional pointer to an {\tt int},
which the library uses to specify the size of the list.

Next it populates the {\tt pingpong\_context} structure with the function
{\tt pp\_init\_ctx}.

The {\tt pingpong\_context} structure wraps all the resources associated with
a connection into one unit.

\lstset{language=C, caption=struct pingpong\_context}
\begin{lstlisting}
59 struct pingpong_context {
60     struct ibv_context  *context;
61     struct ibv_comp_channel *channel;
62     struct ibv_pd       *pd;
63     struct ibv_mr       *mr;
64     struct ibv_cq       *cq;
65     struct ibv_qp       *qp;
66     void            *buf;
67     int          size;
68     int          rx_depth;
69     int          pending;
70     struct ibv_port_attr     portinfo;
71 };
\end{lstlisting}

\lstset{language=C, caption=Initializing the struct pingpong\_context}
\begin{lstlisting}
643 ctx = pp_init_ctx(ib_dev, size, rx_depth, ib_port,
			      use_event, !servername);
\end{lstlisting}

The {\tt ib\_dev} argument is a {\tt struct device *} and comes from {\tt dev\_list}. The argument
{\tt size} specifies the size of the message to be sent (4096 bytes by
default), {\tt rx\_depth}
sets the number of receives to post at a time, {\tt ib\_port} is the port of the HCA and {\tt use\_event}
specifies whether to sleep on CQ events or poll for them.

The function {\tt pp\_init\_ctx} first allocates a buffer of memory, which will be
used to send and receive data. Note that the buffer is {\tt memalign}-ed to
a page, since it is pinned (see section~\ref{sec:Memory Region} for a
definition of pinning).

\lstset{language=C, caption=Allocating a Buffer}
\begin{lstlisting}
320     ctx->buf = memalign(page_size, size);
321     if (!ctx->buf) {
322         fprintf(stderr, "Couldn't allocate work buf.\n");
323         return NULL;
324     }
325
326     memset(ctx->buf, 0x7b + is_server, size);
\end{lstlisting}

Next the {\tt ibv\_context} pointer is populated with a call to
{\tt ibv\_open\_device}. The {\tt ibv\_context} is a structure which
encapsulates information about the device opened for the connection.

\lstset{language=C, caption=Opening a Context}
\begin{lstlisting}
328     ctx->context = ibv_open_device(ib_dev);
329     if (!ctx->context) {
330         fprintf(stderr, "Couldn't get context for %s\n",
331             ibv_get_device_name(ib_dev));
332         return NULL;
333     }
\end{lstlisting}

From the {\tt ``infiniband/verbs.h''} header, the {\tt struct ibv\_context} is as follows:
\lstset{language=C, caption=struct ibv\_context}
\begin{lstlisting}
766 struct ibv_context {
767     struct ibv_device      *device;
768     struct ibv_context_ops  ops;
769     int         cmd_fd;
770     int         async_fd;
771     int         num_comp_vectors;
772     pthread_mutex_t     mutex;
773     void               *abi_compat;
774     struct ibv_more_ops     *more_ops;
775 };
\end{lstlisting}

The {\tt struct ibv\_device *} is a pointer to the device opened for this connection. The
{\tt struct ibv\_context\_ops ops} field contains function pointers to driver specific functions,
which the user need not access directly.

\subsection{Protection Domain}
After the device is opened and the context is created,
the program allocates a protection domain.

\lstset{language=C, caption=Opening a Protection Domain}
\begin{lstlisting}
344     ctx->pd = ibv_alloc_pd(ctx->context);
345     if (!ctx->pd) {
346         fprintf(stderr, "Couldn't allocate PD\n");
347         return NULL;
348     }
\end{lstlisting}

A protection domain, according to the InfiniBand
specification~\cite[page~107]{spec}, allows the client to control which
remote computers can access its memory regions during InfiniBand sends
and receives.

The protection domain mostly exists on the hardware itself. Its user-space
data structure is sparse:

\lstset{language=C, caption=struct ibv\_pd from {\tt ``infiniband/verbs.h''}}
\begin{lstlisting}
308 struct ibv_pd {
311     struct ibv_context     *context;
312     uint32_t        handle;
313 };
\end{lstlisting}

\subsection{Memory Region}
\label{sec:Memory Region}
The {\tt ibv\_rc\_pingpong} program next registers one memory region with the hardware.

When the memory region is registered, two things happen. The memory is
pinned by the kernel, which prevents the physical address from being
swapped to disk. On Linux operating systems, a call to {\tt mlock} is used
to perform this operation. In addition, a translation of the virtual address to the
physical address is given to the HCA.

\lstset{language=C, caption=Registering a Memory Region}
\begin{lstlisting}
350     ctx->mr = ibv_reg_mr(ctx->pd, ctx->buf, size,
			      IBV_ACCESS_LOCAL_WRITE);
351     if (!ctx->mr) {
352         fprintf(stderr, "Couldn't register MR\n");
353         return NULL;
354     }
\end{lstlisting}

The arguments are the protection domain with which to associate the memory region,
the address of the region itself, the size, and the flags. The options for the flags
are defined in {\tt ``infiniband/verbs.h''}.

\lstset{language=C, caption=Access Flags}
\begin{lstlisting}
300 enum ibv_access_flags {
301     IBV_ACCESS_LOCAL_WRITE      = 1,
302     IBV_ACCESS_REMOTE_WRITE     = (1<<1),
303     IBV_ACCESS_REMOTE_READ      = (1<<2),
304     IBV_ACCESS_REMOTE_ATOMIC    = (1<<3),
305     IBV_ACCESS_MW_BIND      = (1<<4)
306 };
\end{lstlisting}

When the memory registration is complete, an {\tt lkey} field
or Local Key is created. According to the InfiniBand Technical
Specification~\cite[Page~76]{spec} the {\tt lkey} is used to identify
the appropriate memory addresses and provide authorization to access them.

\subsection{Completion Queue} The next part of the connection is the
completion queue (CQ), where work completion queue entries are posted.
Please note that you must create the CQ before the QP. As stated
previously, {\tt ibv\_ack\_cq\_events (3)} has helpful examples of how to
manage completion events.

\lstset{language=C, caption=Creating a CQ}
\begin{lstlisting}
356     ctx->cq = ibv_create_cq(ctx->context, rx_depth + 1, NULL,
357                 ctx->channel, 0);
358     if (!ctx->cq) {
359         fprintf(stderr, "Couldn't create CQ\n");
360         return NULL;
361     }
\end{lstlisting}

\subsection{Queue Pairs} Communication in InfiniBand is based on the
concept of queue pairs. Each queue pair contains a send queue and
a receive queue, and must be associated with at least one completion
queue. The queues themselves exist on the HCA. A data structure containing
a reference to the hardware queue pair resources is returned to the user.

First, look at the code to create a QP.

\lstset{language=C, caption=Creating a QP}
\begin{lstlisting}
364         struct ibv_qp_init_attr attr = {
365             .send_cq = ctx->cq,
366             .recv_cq = ctx->cq,
367             .cap     = {
368                 .max_send_wr  = 1,
369                 .max_recv_wr  = rx_depth,
370                 .max_send_sge = 1,
371                 .max_recv_sge = 1
372             },
373             .qp_type = IBV_QPT_RC
374         };
375
376         ctx->qp = ibv_create_qp(ctx->pd, &attr);
377         if (!ctx->qp)  {
378             fprintf(stderr, "Couldn't create QP\n");
379             return NULL;
380         }
\end{lstlisting}

Notice that a data structure which defines the initial attributes of the
QP must be given as an argument.  There are a few other elements in the
data structure, which are optional to define.

The first two elements, {\tt send\_cq} and {\tt recv\_cq},
 associate the QP with a CQ as stated earlier. The
send and receive queue may be associated with the same completion queue.

The cap field points to a {\tt struct ibv\_qp\_cap} and specifies how many send
and receive work requests the queues can hold. The {\tt max\_\{send,recv\}\_sge}
field specifies the maximum number of scatter/gather elements that each work
request will be able to hold. A scatter gather element is used in a direct
memory access (DMA) operation, and each SGE points to a buffer
in memory to be used in the read or write. In this case, the attributes
state that only one buffer may be pointed to at any given time.

The {\tt qp\_type} field specifies what type of connection is to be used,
in this case a reliable connection

Now the queue pair has been created. It must be moved into the
initialized state, which involves a library call. In the initialized
state, the QP will silently drop~\cite[Page~460]{spec} any incoming packets and no work
requests can be posted to the send queue.

\lstset{language=C, caption=Setting QP to INIT}
\begin{lstlisting}
 384         struct ibv_qp_attr attr = {
 385             .qp_state        = IBV_QPS_INIT,
 386             .pkey_index      = 0,
 387             .port_num        = port,
 388             .qp_access_flags = 0
 389         };
 390
 391         if (ibv_modify_qp(ctx->qp, &attr,
 392                   IBV_QP_STATE              |
 393                   IBV_QP_PKEY_INDEX         |
 394                   IBV_QP_PORT               |
 395                   IBV_QP_ACCESS_FLAGS)) {
 396             fprintf(stderr, "Failed to modify QP to INIT\n");
 397             return NULL;
 398         }
\end{lstlisting}

The third argument to {\tt ibv\_modify\_qp} is a bitmask stating which options
should be configured. The flags are specified in {\tt enum ibv\_qp\_attr\_mask}
in {\tt infiniband/verbs.h}.

At this point the {\tt ibv\_rc\_pingpong} program posts a receive work request to the QP.

\lstset{language=C, caption=}
\begin{lstlisting}
650     routs = pp_post_recv(ctx, ctx->rx_depth);
\end{lstlisting}

Look at the definition of {\tt pp\_post\_recv}.

\lstset{language=C, caption=Posting Recv Requests}
\begin{lstlisting}
444 static int pp_post_recv(struct pingpong_context *ctx, int n)
445 {
446     struct ibv_sge list = {
447         .addr   = (uintptr_t) ctx->buf,
448         .length = ctx->size,
449         .lkey   = ctx->mr->lkey
450     };
451     struct ibv_recv_wr wr = {
452         .wr_id      = PINGPONG_RECV_WRID,
453         .sg_list    = &list,
454         .num_sge    = 1,
455     };
456     struct ibv_recv_wr *bad_wr;
457     int i;
458
459     for (i = 0; i < n; ++i)
460         if (ibv_post_recv(ctx->qp, &wr, &bad_wr))
461             break;
462
463     return i;
464 }
\end{lstlisting}

The {\tt ibv\_sge} list is the list pointing to the scatter/gather elements
(in this case, a list of size~1).  To review, the SGE is a pointer to
a memory region which the HCA can read to or write from.

Next is the {\tt ibv\_recv\_wr structure}. The first field, {\tt wr\_id}, is a field
set by the program to identify the work request.  This is needed
when checking the completion queue elements; it specifies which work
request completed.

The work request given to {\tt ibv\_post\_recv} is actually a linked
list, of length 1.

\lstset{language=C, caption=Linked List}
\begin{lstlisting}
451     struct ibv_recv_wr wr = {
452         .wr_id      = PINGPONG_RECV_WRID,
453         .sg_list    = &list,
454         .num_sge    = 1,
455     };
\end{lstlisting}

If one of the work requests fails, the library
will set the {\tt bad\_wr} pointer to the failed {\tt wr} in the linked list.

{\bf Receive buffers must be posted before any sends.} It is common
practice to loop over the {\tt ibv\_post\_recv} call to post numerous buffers
at the beginning of execution. Eventually these buffers will be used up;
internal flow control must be implemented by the applications to ensure
that sends are not posted without corresponding receives.

\subsection{Connecting}

The next step occurs in {\tt pp\_client\_exch\_dest}
and {\tt pp\_server\_exch\_dest}. The QPs need to be configured to point to a
matching QP on a remote node. However, the QPs currently have no means of
locating each other. The processes open an out-of-band TCP socket
and transmit the needed information. That information, once manually
communicated, is given to the driver and then each side's QP is configured to
point at the other. (The OFED {\tt librdma\_cm} library is an alternative
to explicit out-of-band TCP.)

So what information needs to be exchanged/configured? Mainly the LID, QPN,
and PSN. The LID is the ``Local Identifier'' and it is a unique number given
to each port when it becomes active. The QPN is the Queue Pair Number,
and it is the identifier assigned to each queue on the HCA. This is
used to specify to what queue messages should be sent. Finally, the
destinations must share their PSNs.

The PSN stands for Packet Sequence Number. In a reliable connection it
is used by the HCA to verify that packets are coming in order and that
packets are not missing. The initial PSN, for the first packet, must be
specified by the user code. If it is too similar to a recently used PSN,
the hardware will assume that the incoming packets are stale packets
from an old connection and reject them.

The GID, seen in the code sample below, is a 128-bit unicast or multicast
identifier used to identify an endport~\cite[page~74]{spec}. The link layer specifies
which interconnect the software is running on; there are other interconnects
that OFED supports, though that is not within the scope of this paper.

Within {\tt pp\_connect\_ctx} the information, once transmitted, is used to
connect the QPs into an end-to-end context.

\lstset{language=C, caption=Setting Up Destination Information}
\begin{lstlisting}
665     my_dest.lid = ctx->portinfo.lid;
666     if (ctx->portinfo.link_layer == IBV_LINK_LAYER_INFINIBAND &&
							!my_dest.lid) {
667         fprintf(stderr, "Couldn't get local LID\n");
668         return 1;
669     }
670
671     if (gidx >= 0) {
672         if (ibv_query_gid(ctx->context, ib_port, gidx, &my_dest.gid)) {
673             fprintf(stderr, "Could not get local gid for gid index "
							      "%d\n", gidx);
674             return 1;
675         }
676     } else
677         memset(&my_dest.gid, 0, sizeof my_dest.gid);
678
679     my_dest.qpn = ctx->qp->qp_num;
680     my_dest.psn = lrand48() & 0xffffff;
\end{lstlisting}

\begin{figure}
 \centering
 \includegraphics[width=300pt,keepaspectratio=true]{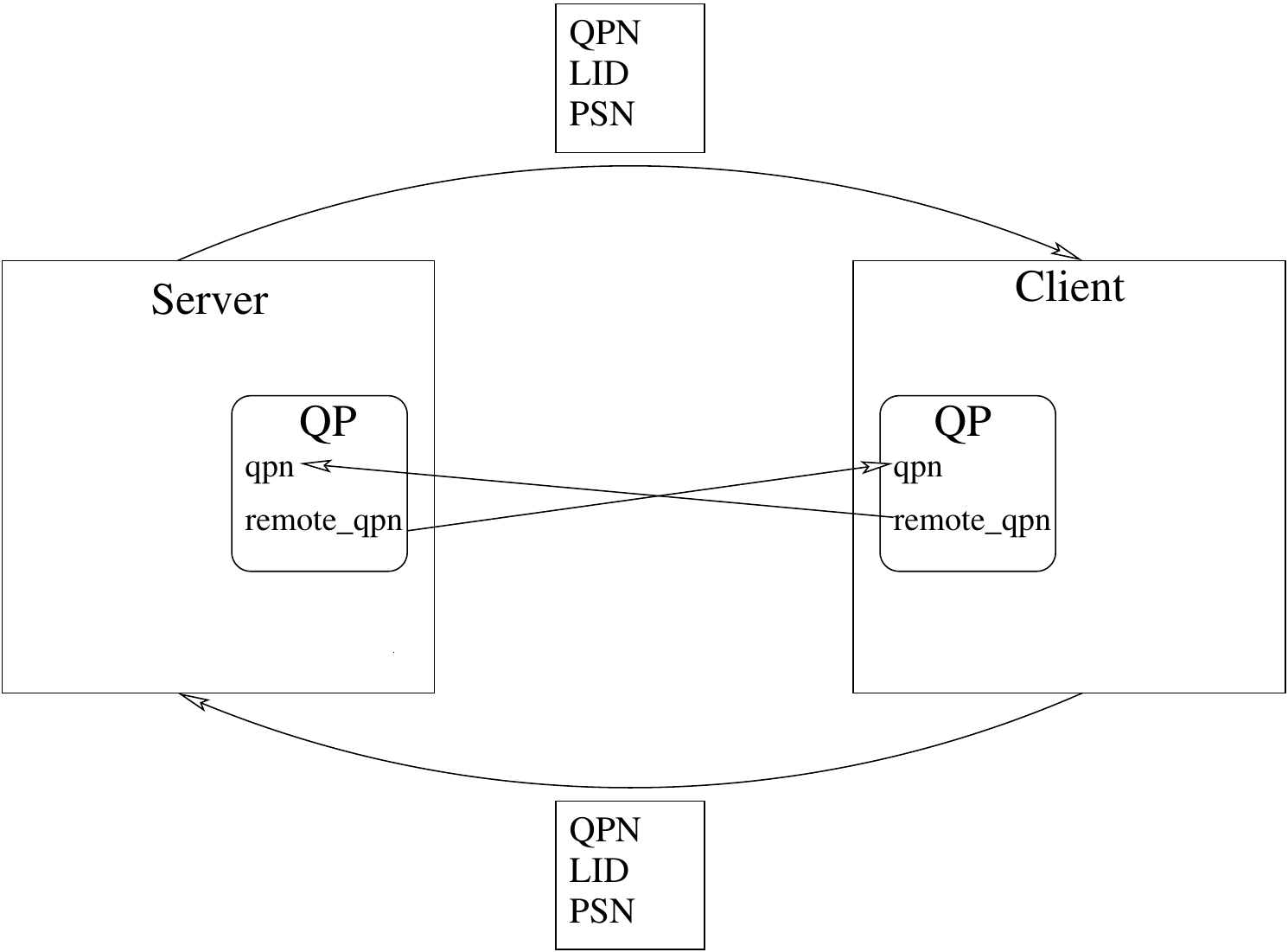}
 \caption{Exchanging QPNs via TCP}
 \label{fig:TCP}
\end{figure}

The {\tt my\_dest} data structure is filled and then transmitted via TCP.
Figure~\ref{fig:TCP} illustrates this data transfer.

\subsubsection{Modifying QPs}
Look at the attributes given to the {\tt ibv\_modify\_qp} call.

\lstset{language=C, caption=Moving QP to Ready to Recv}
\begin{lstlisting}
84     struct ibv_qp_attr attr = {
85         .qp_state       = IBV_QPS_RTR,
86         .path_mtu       = mtu,
87         .dest_qp_num        = dest->qpn,
88         .rq_psn         = dest->psn,
89         .max_dest_rd_atomic = 1,
90         .min_rnr_timer      = 12,
91         .ah_attr        = {
92             .is_global  = 0,
93             .dlid       = dest->lid,
94             .sl     = sl,
95             .src_path_bits  = 0,
96             .port_num   = port
97         }
98     };

106     if (ibv_modify_qp(ctx->qp, &attr,
107               IBV_QP_STATE              |
108               IBV_QP_AV                 |
109               IBV_QP_PATH_MTU           |
110               IBV_QP_DEST_QPN           |
111               IBV_QP_RQ_PSN             |
112               IBV_QP_MAX_DEST_RD_ATOMIC |
113               IBV_QP_MIN_RNR_TIMER)) {
114         fprintf(stderr, "Failed to modify QP to RTR\n");
115         return 1;
116     }
\end{lstlisting}

As you can see {\tt .qp\_state} is set to {\tt IBV\_QPS\_RTR}, or
Ready-To-Receive. The three fields swapped over TCP, the PSN, QPN, and
LID, are now given to the hardware. With this information, the QPs are
registered with each other by the hardware, but are not ready to begin
exchanging messages. The {\tt min\_rnr\_timer} is the time, in seconds,
between retries before a timeout occurs.

The QP must be moved into the Ready-To-Send state before the ``connection'' process is complete.

\lstset{language=C, caption=Moving QP to Ready to Send}
\begin{lstlisting}
118     attr.qp_state       = IBV_QPS_RTS;
119     attr.timeout        = 14;
120     attr.retry_cnt      = 7;
121     attr.rnr_retry      = 7;
122     attr.sq_psn     = my_psn;
123     attr.max_rd_atomic  = 1;
124     if (ibv_modify_qp(ctx->qp, &attr,
125               IBV_QP_STATE              |
126               IBV_QP_TIMEOUT            |
127               IBV_QP_RETRY_CNT          |
128               IBV_QP_RNR_RETRY          |
129               IBV_QP_SQ_PSN             |
130               IBV_QP_MAX_QP_RD_ATOMIC)) {
131         fprintf(stderr, "Failed to modify QP to RTS\n");
132         return 1;
133     }
\end{lstlisting}

The {\tt attr} used to move the QP into {\tt IBV\_QPS\_RTS} is the same
{\tt attr} used in the previous call. There is no need to zero out the
structure because the bitmask, given as the third argument, specifies which
fields should be set.

After the QP is moved into the Ready-To-Send state, the connection
(end-to-end context) is ready.

\subsection{Sending Data}
Since the server already posted receive buffers, the client will now
post a ``send'' work request.

\lstset{language=C, caption=Client Posting Send}
\begin{lstlisting}
468     struct ibv_sge list = {
469         .addr   = (uintptr_t) ctx->buf,
470         .length = ctx->size,
471         .lkey   = ctx->mr->lkey
472     };
473     struct ibv_send_wr wr = {
474         .wr_id      = PINGPONG_SEND_WRID,
475         .sg_list    = &list,
476         .num_sge    = 1,
477         .opcode     = IBV_WR_SEND,
478         .send_flags = IBV_SEND_SIGNALED,
479     };
480     struct ibv_send_wr *bad_wr;
481
482     int rslt = ibv_post_send(ctx->qp, &wr, &bad_wr);
\end{lstlisting}

The {\tt wr\_id} is an ID specified by the programmer to identify the
completion notification corresponding with this work request. In addition,
the flag {\tt IBV\_SEND\_SIGNALED} sets the completion notification
indicator. According to {\tt ibv\_post\_send (3)}, it is only relevant
if the QP is created with {\tt sq\_sig\_all = 0}.

\subsection{Flow Control}

Programmers must implement their own flow control when working
with the verbs API. Let us examine the flow control used in {\tt
ibv\_rc\_pingpong}. Remember from earlier that a client cannot post a send
if its remote node does not have a buffer waiting to receive the data.

Flow control must be used to ensure that receivers do not exhaust their
supply of posted receives. Furthermore, the CQ must not overflow. If the
client does not pull CQEs off the queue fast enough, the CQ is thrown
into an error state, and can no longer be used.

You can see at the top of the loop, which will send/recv the data,
that {\tt ibv\_rc\_pingpong} tracks the send and recv count.

\lstset{language=C, caption=Flow Control}
\begin{lstlisting}
717     rcnt = scnt = 0;
718     while (rcnt < iters || scnt < iters) {
\end{lstlisting}

Now the code will poll the CQ for two completions; a send completion and a receive completion.

\lstset{language=C, caption=Polling the CQ}
\begin{lstlisting}
745            do {
746                 ne = ibv_poll_cq(ctx->cq, 2, wc);
747                 if (ne < 0) {
748                     fprintf(stderr, "poll CQ failed %d\n", ne);
749                     return 1;
750                 }
751
752            } while (!use_event && ne < 1);
\end{lstlisting}

The {\tt use\_event} variable specifies whether or not the program
should sleep on CQ events. By default, {\tt ibv\_rc\_pingpong} will
poll. Hence the while-loop. On success, {\tt ibv\_poll\_cq} returns the number
of completions found.

Next, the program must account for how many sends and receives have been posted.

\lstset{language=C, caption=Flow Control Accounting}
\begin{lstlisting}
762         switch ((int) wc[i].wr_id) {
763         case PINGPONG_SEND_WRID:
764             ++scnt;
765             break;
766
767         case PINGPONG_RECV_WRID:
768             if (--routs <= 1) {
769                 routs += pp_post_recv(ctx, ctx->rx_depth - routs);
770                 if (routs < ctx->rx_depth) {
771                     fprintf(stderr,
772                         "Couldn't post receive (%d)\n",
773                         routs);
774                     return 1;
775                 }
776             }
777
778             ++rcnt;
779             break;
780
781         default:
782             fprintf(stderr, "Completion for unknown wr_id %d\n",
783                     (int) wc[i].wr_id);
784             return 1;
785         }
\end{lstlisting}

The ID given to the work request is also given to its associated work
completion, so that the client knows what WQE the CQE is associated
with. In this case, if it finds a completion for a send event, it
increments the send counter and moves on.

The case for {\tt PINGPONG\_RECV\_WRID} is more interesting, because
it must make sure that receive buffers are always available. In this case
the {\tt routs} variable indicates how many recv buffers are available.
So if only one buffer remains available, {\tt ibv\_rc\_pingpong} will post
more recv buffers. In this case, it calls {\tt pp\_post\_recv} again,
which will post another 500 (by default).  After that it increments the
recv counter.

Finally, if more sends need to be posted, the program will post another
send before continuing the loop.

\lstset{language=C, caption=Posting Another Send}
\begin{lstlisting}
787                 ctx->pending &= ~(int) wc[i].wr_id;
788                 if (scnt < iters && !ctx->pending) {
789                     if (pp_post_send(ctx)) {
790                         fprintf(stderr, "Couldn't post send\n");
791                         return 1;
792                     }
793                     ctx->pending = PINGPONG_RECV_WRID |
794                                PINGPONG_SEND_WRID;
795                 }
\end{lstlisting}

\subsection{ACK}
The {\tt ibv\_rc\_pingpong} program will now ack the completion events with a call to
\\ {\tt ibv\_ack\_cq\_events}. To avoid races, the CQ destroy operation will wait for all
completion events returned by {\tt ibv\_get\_cq\_event} to be acknowledged.
The call to {\tt ibv\_ack\_cq\_events} must take a mutex internally, so it is best
to ack multiple events at once.

\lstset{language=C, caption=}
\begin{lstlisting}
816     ibv_ack_cq_events(ctx->cq, num_cq_events);
\end{lstlisting}

As a reminder, {\tt ibv\_ack\_cq\_events (3)} has helpful sample code.

\section{Conclusion} InfiniBand is the growing standard for supercomputer
interconnects, even appearing in departmental clusters.  The API is
complicated and sparsely documented, and the sample program provided
by OFED, {\tt ibv\_rc\_pingpong}, does not fully explain the
functionality of the verbs. This paper will hopefully enable the reader
to better understand the verbs interface.

\section{Acknowledgements} Gene Cooperman (Northeastern University)
and Jeff Squyres (Cisco) contributed substantially to the organization,
structure, and content of this document. Jeff also took the time
to discuss the details of InfiniBand with me. Josh Hursey (Oak Ridge
National Laboratory) shared his knowledge of InfiniBand with me along
the way. Roland Dreier (PureStorage) pointed out, and corrected, a mistake
in my explanation of acks.

\bibliographystyle{plain}

\end{document}